\documentclass[journal]{IEEEtran}

\usepackage[dvips]{graphicx} 
\usepackage[cmex10]{amsmath}
\usepackage{fixltx2e} 
\usepackage{stfloats} 

\begin{document}
\title{Finite-difference time-domain formulation of stochastic noise in 
  macroscopic atomic systems}
\author{Jonathan Andreasen, \IEEEmembership{Student Member, IEEE,}
  Hui Cao, \IEEEmembership{Fellow, OSA}%
  \thanks{The authors are with the Department of Applied Physics, 
    Yale University, New Haven, CT 06520 USA and J. Andreasen is 
    also with the the Department of 
    Physics and Astronomy, Northwestern University, Evanston, IL 60208 USA
    (email: hui.cao@yale.edu; j-andreasen@northwestern.edu).}}
\maketitle

\begin{abstract}
  A numerical model based on the finite-difference time-domain method is
  developed to simulate fluctuations which accompany the dephasing of atomic 
  polarization and the decay of excited state's population.
  This model is based on the Maxwell-Bloch equations with \textit{c}-number 
  stochastic noise terms.
  We successfully apply our method to a numerical simulation of the atomic 
  superfluorescence process.
  This method opens the door to further studies of the effects of stochastic 
  noise on light-matter interaction and transient processes in complex systems 
  without prior knowledge of modes.
\end{abstract}

\begin{IEEEkeywords}
  Noise,
  spontaneous emission, 
  stochastic processes,
  FDTD methods, 
  Maxwell equations
\end{IEEEkeywords}

\section{Introduction}
\IEEEPARstart{T}{he} finite-difference time-domain (FDTD) method \cite{tafl05}
has been extensively used in solving Maxwell's equations for dynamic 
electromagnetic (EM) fields. 
The incorporation of auxiliary differential equations, such as the rate 
equations for atomic populations \cite{nagra98} and the Bloch equations 
for the density of states of atoms \cite{ziol95}, has lead to comprehensive 
studies of light-matter interaction. 
Although the FDTD method has become a powerful tool in computational 
electrodynamics, it has been applied mostly to classical or semiclassical 
problems without noise. 

Noise plays an important role in light-matter interaction. 
Marcuse solved 
the rate equations for light intensity and electron population including noise 
terms \cite{marc84I} to illustrate the effect of noise on lasing mode 
dynamics \cite{marc84II}.
Gray and Roy extended the formulation by adding noise to the field equation 
in order to study the laser line shape \cite{gray89}.
Starting from a microscopic Hamiltonian, 
Kira \textit{et al.} developed a semiconductor theory including spontaneous 
emission to describe semiconductor lasers \cite{kira99}.
While considerable progress has been made, these models remain in the modal
picture. 
Knowledge of mode properties is required to characterize the noise, making it 
difficult to study complex systems in which the mode information is unknown 
\textit{a priori}. 
Without invoking the modal picture, Hofmann and Hess obtained the quantum 
Maxwell-Bloch equations including spatiotemporal fluctuations \cite{hofmann99}. 
Although it was useful to study spatial and temporal coherence in diode lasers, 
this formalism was based on the assumption that the temporal fluctuations of 
carrier density and photon density were statistically independent, which 
often broke down above the lasing threshold.
A FDTD simulation of microcavity lasers including quantum fluctuations was 
also done recently \cite{slav04}. 
This simplified model added white Gaussian noise as a source to the 
electric field. 
The noise amplitude depended only on the excited state's lifetime. 
The dephasing process, which was much faster than the excited state's 
population decay, should have induced more noise but was neglected. 

Our goal is to develop a FDTD-based numerical method to simulate fluctuations 
in macroscopic systems caused by interactions of atoms and photons with 
reservoirs (heatbaths). 
Such interactions induce temporal decay of photon number, atomic polarization 
and excited state's population, which can be described phenomenologically 
by decay constants. 
The fluctuation-dissipation theorem demands temporal fluctuations or noise to 
accompany these decays. 
We intend to incorporate such noise in a way compatible with the FDTD method, 
that allows one to study the light-matter interaction in complex systems 
without prior knowledge of modes.
In a previous work \cite{Andreasen08}, we included noise caused by the 
interaction of light field with external reservoir in an open system. 
In this paper, we develop a numerical model to simulate noise caused by the 
interaction of atoms with reservoirs such as lattice vibrations and atomic 
collisions.
As an example, we apply the method to a numerical simulation of 
superfluorescence in a macroscopic system where the dominant noise is from 
the atoms rather than the light field.

We start with the Bloch equations for two-level atoms in one dimension (1D) 
where the direction of light propagation is along the \textit{x}-axis.
\setlength{\arraycolsep}{0.0em}
\begin{eqnarray}
  \left (
  \begin{array}{c}
    \dot{\rho}_1\\
    \dot{\rho}_2\\
    \dot{\rho}_3
  \end{array}
  \right )&{}={}&
  \left (
  \begin{array}{ccc}
    0 & \omega_0 & 0 \\
    -\omega_0 & 0 & 2\Omega_R \\
    0 & -2\Omega_R & 0
  \end{array}
  \right )
  \left (
  \begin{array}{c}
    \rho_1\\
    \rho_2\\
    \rho_3
  \end{array}
  \right ) \nonumber\\
  &&{-} 
  \left (
  \begin{array}{ccc}
    1/T_2 & 0 & 0 \\
    0 & 1/T_2 & 0 \\
    0 & 0 & 1/T_1 + P_r
  \end{array}
  \right) 
  \left(
  \begin{array}{c}
    \rho_1 \\
    \rho_2 \\
    \rho_3 - \rho_{3}^{(s)}
  \end{array}
  \right )\label{eq:blochvectordot},
\end{eqnarray}
where $\Omega_R \equiv \gamma E_z/\hbar$ is the Rabi frequency, 
$\omega_0$ the atomic transition frequency, 
$E_z$ the electric field which is parallel to the 
\textit{z}-axis, $\gamma$ the dipole coupling term. 
Phenomenological decay times due to decoherence $T_2$ and the excited state's 
lifetime $T_1$ (which includes spontaneous emission and non-radiative 
recombination) are appended. 
In the absence of strong light confinement, which holds for macroscopic 
systems, $T_1$ and $T_2$ can be considered independent of the local density 
of states (LDOS). Hence, they do not have a dependence on spatial location nor frequency. 
We also include incoherent pumping of atoms from level 1 to level 2. 
The rate is proportional to the population in level 1, and can be written as $P_r \rho_{11}$.
$\rho_{3}^{(s)}$ represents the steady-state value of $\rho_3$ when $E_z = 0$.

The relations between the Bloch vector and the density matrix are 
\begin{eqnarray}
  \rho_1 =& \rho_{12} + \rho_{21}\nonumber\\
  \rho_2 =& i(\rho_{12} - \rho_{21})\nonumber\\
  \rho_3 =& \rho_{22} - \rho_{11}.\label{eq:blochvec}
\end{eqnarray}
The total polarization $P_z$ of $N$ atoms in a volume $V$ is 
$P_z = -(N/V)|\gamma|\rho_1$
and inserted into the Maxwell's equations
\setlength{\arraycolsep}{0.0em}
\begin{eqnarray}
  \frac{dH_y}{dt}&{}={}& -\frac{1}{\mu_0}\frac{dE_z}{dx} \nonumber\\
  \frac{dE_z}{dt}&{}={}& \frac{1}{\epsilon}\frac{dH_y}{dx} - 
  \frac{1}{\epsilon}\frac{dP_z}{dt}.
\end{eqnarray}
\setlength{\arraycolsep}{5pt}

The atom-reservoir interactions not only cause decay of the Bloch vector, 
but also introduce noise according to the fluctuation-dissipation 
theorem.
In Section \ref{sec:model}, we describe the model developed to include 
noise in the Maxwell-Bloch equations. 
The FDTD implementation of this model is presented in Section 
\ref{sec:implementation}.
In Section \ref{sec:results}, we simulate atomic superfluorescence 
and compare the results to previous experimental data and 
quantum-mechanical calculations. 

\section{Noise Model\label{sec:model}}
Starting from the quantum Langevin equation within the Markovian approximation, 
Drummond and Raymer derived a set of stochastic \textit{c}-number differential 
equations describing light propagation and atom-light interaction in the 
many-atom limit \cite{drum91}. 
The noise sources in these equations are from both the damping and the 
nonlinearity in the Hamiltonian. 
The latter represents the nonclassical component of noise, giving rise to 
nonclassical statistical behavior. 
Since our primary interests lie with classical behavior of macroscopic systems, 
such as superfluorescence and lasing, we neglect the nonclassical noise in 
this paper. 
The amplitude of classical noise accompanying the field decay is proportional to $\sqrt{n}$, where $n$ is the thermal photon number. 
At room temperature the number of thermal photons at visible frequencies ($\hbar \omega \sim$ 1 eV) is on the order of $10^{-17}$.  
This can be interpreted in a quantum mechanical picture as that most of the time there are no thermal photons at visible frequencies in the system.
Thus, the noise related to field decay is neglected in this paper. 
At higher temperatures or longer wavelengths, this noise becomes significant
and it can be incorporated into the FDTD algorithm following the approach we developed in our previous work \cite{Andreasen08}.

The classical noise related to the pumping and decay of the atomic density 
matrix can be expressed as
\setlength{\arraycolsep}{0.0em}
\begin{eqnarray}
  \Gamma_{12}&{}={}& (\xi_1 + i\xi_2)\sqrt{\gamma_p \rho_{22}}\nonumber\\
  \Gamma_{21}&{}={}& (\xi_1 - i\xi_2)\sqrt{\gamma_p \rho_{22}}\nonumber\\
  \Gamma_{22}&{}={}& \xi_3\sqrt{\rho_{22}/T_1+ P_r \rho_{11})}.\label{eq:Gamma122122}
\end{eqnarray}
These noise terms are associated with $\rho_{12}$, $\rho_{21}$, and $\rho_{22}$
respectively.
$\gamma_p = 1/T_2 - 1/2T_1$.  
The $\xi_j$ terms are real, Gaussian, random variables with zero mean and the 
following correlation relation
\begin{equation}
  \left< \xi_j(t)\xi_k(t')\right> = \delta_{jk}\delta(t-t'),
\end{equation}
where $j, k = 1, 2, 3$. 
The noise terms $\Gamma_{12}$ and $\Gamma_{21}$ represent fluctuations 
corresponding to decoherence by dephasing, while $\Gamma_{22}$ is the 
fluctuation corresponding to relaxation of 
and pumping to 
the excited state's population.  
Only the linear term for pump noise is included here, a common first 
order approximation \cite{haken83}.
Furthermore, because we assume $T_2 \ll T_1$, pump fluctuations are 
neglected in $\Gamma_{12}$ and $\Gamma_{21}$ since they are orders of magnitude 
smaller than noise due to dephasing.
According to (\ref{eq:blochvec}), the noise terms for the Bloch vector are 
reduced to real variables as
\setlength{\arraycolsep}{0.0em}
\begin{eqnarray}
  \Gamma_1 &{}={}& 2\xi_1\sqrt{\gamma_p\rho_{22}}\nonumber\\
  \Gamma_2 &{}={}& -2\xi_2\sqrt{\gamma_p\rho_{22}}\nonumber\\
  \Gamma_3 &{}={}& 2\xi_3\sqrt{\rho_{22}/T_1 + P_r \rho_{11})}. \label{eq:Gamma1222}
\end{eqnarray}
\setlength{\arraycolsep}{5pt}
They can be added directly to (\ref{eq:blochvectordot}).

In a 1D system, the total number of atoms $N$ are split equally among $M$ 
grid cells, giving the number of atoms per cell $N_s=N/M$.
All quantities are defined at each individual grid cell, \textit{e.g.}
the term $\rho_3(x)$ is the \textit{number} of inverted atoms in one cell at 
position $x$.
The number of atoms in each cell is assumed to be constant assuring 
$\dot{\rho}_{11}+\dot{\rho}_{22}=0$.
We forcibly keep $N_s$ constant via the relation $\rho_{11} = N_s - \rho_{22}$ 
and only calculate the excited state's population $\rho_{22}(t)$. 
The final stochastic equations to be solved are
\setlength{\arraycolsep}{0.0em}
\begin{eqnarray}
  \frac{d\rho_1(x,t)}{dt} &{}={}& \omega_0\rho_2(x,t) - \frac{1}{T_2}\rho_1(x,t) + 
  \Gamma_1(x,t) \nonumber\\
  \frac{d\rho_2(x,t)}{dt} &{}={}& -\omega_0\rho_1(x,t) + 
  \frac{2|\gamma|}{\hbar}E_z(x,t)\left(2\rho_{22}(x,t)-N_s\right)\nonumber\\
  &&{-} \frac{1}{T_2}\rho_2(x,t) +   \Gamma_2(x,t) \nonumber\\
  \frac{d\rho_{22}(x,t)}{dt} &{}={}& -\frac{|\gamma|}{\hbar}E_z(x,t)\rho_2(x,t)
  - \frac{1}{T_1}\rho_{22}(x,t)
  \nonumber\\
  &&{+} P_r(N_s-\rho_{22}(x,t)) + \Gamma_{22}(x,t). \label{eq:theory}
\end{eqnarray}
\setlength{\arraycolsep}{5pt}
In the above equation, the steady-state value of $\rho_3$ in (\ref{eq:blochvectordot}) is substituted by $\rho_3^{(s)} = N_s (T_1 P_r - 1) / (T_1 P_r + 1)$, 
an expression obtained by setting the time derivatives in (\ref{eq:blochvectordot}) to zero. 
$\rho_{11}$ in the expression of $\Gamma_{22}$ in (\ref{eq:Gamma122122}) can be replaced by $N_s - \rho_{22}$. 

\section{Numerical Implementation\label{sec:implementation}}
The most commonly used method of solving the Maxwell-Bloch equations 
is the ``strongly coupled method.''
With $\Delta t$ being the time step, $E$ and $\rho$ are both computed at 
$n\Delta t$, $(n+1)\Delta t$, \textit{etc.}, while $H$ is computed at 
$(n-1/2)\Delta t$, $(n+1/2)\Delta t$, \textit{etc.} 
This produces equations with coupled terms such as $E^{n+1}\rho^{n+1}$ that 
must be solved by a predictor-corrector scheme (as used in \cite{ziol95}) or 
a fixed-point procedure, both of which are computationally inefficient.
Therefore, we use a weakly coupled method that is easily implemented and 
efficient for 1D systems.

The weakly coupled method was put forth by Bid\'egaray \cite{bid03}.
The electric field $E_z$ is computed at times $n\Delta t$, $(n+1)\Delta t$, 
but $\rho$ is calculated at $(n-1/2)\Delta t$, $(n+1/2)\Delta t$, thereby 
decoupling those discretized equations and creating a simple leap-frog 
type propagation system for 1D.
The noise terms in (\ref{eq:theory}) are present throughout the entirety 
of the simulation and thus, should be incorporated efficiently.
After discretization, the $\xi_i$ terms are correlated according to
$\left< \xi_j(x_u,t_m)\xi_k(x_v,t_n)\right> = 
(1/\Delta t)\delta_{jk} \delta_{uv} \delta_{mn}$,
and can be generated quickly with the Marsaglia and Bray modification 
of the Box-M\"{u}ller Transformation \cite{brys91}.
Because the noise terms contain $\sqrt{\rho_{22}}$, as seen  
in (\ref{eq:Gamma122122}) and (\ref{eq:Gamma1222}), we are not able to use 
the weakly coupled scheme to solve for $\rho_1,\rho_2$ and $\rho_{22}$ as 
precisely as possible.
Instead, the approximation of using the previous time step value 
$\sqrt{\rho_{22}^{n-1/2}}$ is employed.
It is valid as long as the atomic population is varying slowly. 
For the simulation of superfluorescence in Sec. \ref{sec:results}, the  
\textit{maximum} change of $\rho_{22}$ over one time step $\Delta t$ is only 0.0007\%.  

\begin{figure*}
  \centering
  \includegraphics[width=16cm]{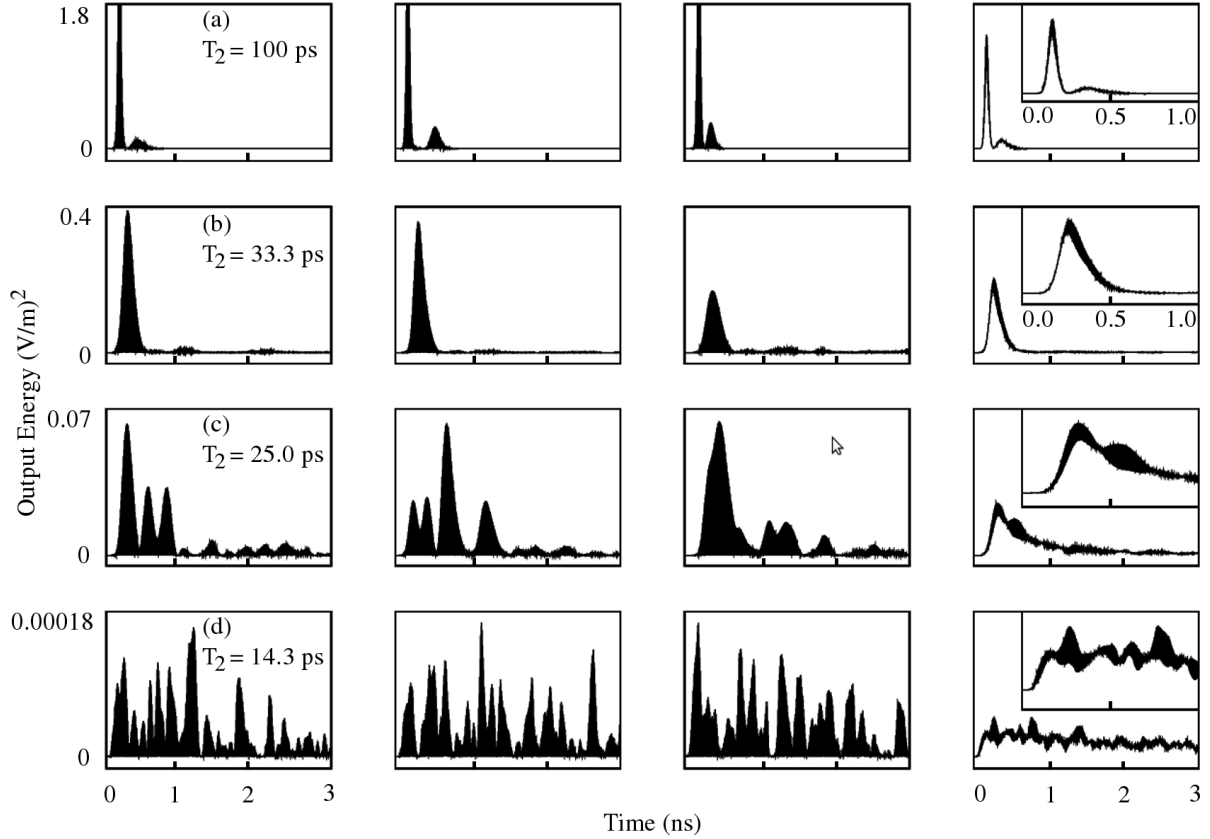}
  \caption{\label{fig:fig1}
    Numerical results of the output EM energy 
    from initially-inverted two-level atoms, 
    obtained by FDTD solution of the Maxwell-Bloch equations with noise.
    The left three columns show the output energy for three 
    random realizations. 
    The last column on the right shows the output energy averaged over 
    30 random realizations.
    All insets in the last column magnify the temporal range $0$ $< t < 1$ ns.
    Dephasing time $T_2$ = 100 ps (first row), 33.3 ps (second row), 
    25.0 ps (third row), and 14.3 ps (fourth row). 
  }
\end{figure*}

The discretized equations with noise are 
\setlength{\arraycolsep}{0.0em}
\begin{subequations}
\begin{eqnarray}
  E_z^{n+1} &{}={}& E_z^n + \frac{\Delta t}{\epsilon}\frac{dH_y}{dx} - 
  \Delta tA\rho_1^{n+1/2} \\ \nonumber
  &&{+} \Delta tB\rho_2^{n+1/2} \\
  H_y^{n+1/2} &{}={}& H_y^{n-1/2} - \frac{\Delta t}{\mu_0}\frac{dE_z}{dx}\\
  \rho_1^{n+1/2} &{}={}& \rho_1^{n-1/2} + \frac{1}{2}\Delta t\omega_0
  \left(\rho_2^{n+1/2} + \rho_2^{n-1/2}\right) \nonumber\\
  &&{-} \frac{1}{2}\frac{\Delta t}{T_2}\left(\rho_1^{n+1/2}+\rho_1^{n-1/2}\right)
  + \Delta t\Gamma_1 \\
  \rho_2^{n+1/2} &{}={}& \rho_2^{n-1/2} - 
  \frac{1}{2}\Delta t\omega_0\left(\rho_1^{n+1/2}+\rho_1^{n-1/2}\right) 
  \nonumber\\
  &&{+} \frac{2\Delta t|\gamma|}{\hbar}E_z^n
  \left(\rho_{22}^{n+1/2}+\rho_{22}^{n-1/2}-N_s\right)\nonumber\\
  &&{-}\frac{1}{2}\frac{\Delta t}{T_2}\left(\rho_2^{n+1/2}+\rho_2^{n-1/2}\right)
  + \Delta t\Gamma_2\\
  \rho_{22}^{n+1/2} &{}={}& \rho_{22}^{n-1/2} - 
  \frac{1}{2}\frac{\Delta t|\gamma|}{\hbar}E_z^n
  \left(\rho_2^{n+1/2}+\rho_2^{n-1/2}\right) 
  \nonumber\\
  &&{-} \frac{1}{2} \Delta t \left(\frac{1}{T_1} + P_r\right)
  \left(\rho_{22}^{n+1/2}+\rho_{22}^{n-1/2}\right)
  \nonumber\\
  &&{+} \Delta t P_r N_s
  + \Delta t\Gamma_{22},
\end{eqnarray}
\label{eq:numfinal}
\end{subequations}
where we have defined
$A \equiv {|\gamma|}/{V_s\epsilon T_2}$ and 
$B \equiv {|\gamma| \omega_0}/{V_s\epsilon}$.
These equations are solved to obtain the final FDTD equations for 
$E_z$, $H_y$, $\rho_1$, $\rho_2$ and $\rho_{22}$.

\section{Results and Discussion\label{sec:results}}

We apply the Maxwell-Bloch equations with noise to a FDTD simulation of 
superfluorescence (SF) and compare the results to previous data obtained 
experimentally \cite{malcuit87} and theoretically \cite{maki89}.
SF is the cooperative radiation of an initially inverted but incoherent 
two-level medium resulting from spontaneous buildup of a macroscopic 
coherent dipole. 
This is an interesting and suitable case to study with our method because 
both spatial propagation of light and noise are important. 
Noise caused by collisional dephasing can seriously disturb SF and change 
the emission character to amplified spontaneous emission (ASE). 
We simulate the transition from SF to ASE with increasing dephasing rate, 
corresponding to the experiment by Malcuit {\it et al.} on super-oxide 
ions in potassium chloride (KCl:O$_2^-$) \cite{malcuit87}. 

Experimentally the ions inside a cylinder of diameter $d$ = 80 $\mu$m and 
length $L$ = 7 mm were excited by a short pulse. 
The total number of excited ions is $N = 3 \times 10^9$. 
The emission wavelength is $\lambda=629$ nm. 
The Fresnel number for the excitation cylinder is $F = A / \lambda L \sim 1$, 
where $A$ is the area of the cylinder cross-section. 
$T_1 = 76$ ns, and $T_2$ was varied via temperature change. 
The ``cooperative lifetime'' or the duration of SF pulse
$\tau_r={8\pi A T_1}/{3\lambda^2 N}$ is 2.7 ps.  
The estimated delay time for the SF peak after the excitation pulse
\begin{equation}
  \tau_d=\tau_r\left[\frac{1}{4}\ln(2\pi N)\right]^2\label{eq:taud}
\end{equation}
is 94 ps.  

Since $F \sim 1$, the EM modes
propagating non-parallel to the cylinder axis are not supported \cite{haake79}. 
Those modes propagating along the cylinder axis do not have a strong radial 
dependence, nor are there significant diffraction losses. 
Thus the system can be considered as 1D in our FDTD simulation. 
The grid resolution is $\Delta x=70$ nm and the total running time is
$\tau_{sim}=3$ ns. 
The Courant number $S$ is set to 0.999999. 
The magic time step, $S=1$, was seen to cause an instability in some cases.
The value $S=1/2$, however, does not propagate the large sudden impulses of 
the noise accurately.
Setting $S=1 - 10^{-6}$ preserves the accuracy to an acceptable degree 
while eliminating the instability at $S=1$.
There is some numerical dispersion and reflection from the absorbing 
boundary layer, but the error is of the order $10^{-6}$. 
Ignoring non-radiative recombination, the atomic dipole coupling term
$|\gamma| = \sqrt{{3\lambda^3\hbar\epsilon_0}/ {8\pi^2T_1}} 
= 1.1 \times 10^{-29}$ C$\cdot$m. 
  
The simulation is started with the initial condition of all the atoms being 
excited ($\rho_{22}=N_s$).
However, because the atomic population and polarization operators do not 
commute, the uncertainty principle demands a nonvanishing variance in the 
initial values of the Bloch vector \cite{haake79}.
This results in a tipping angle $\theta$ of the initial Bloch vector 
away from the top of Bloch sphere ($\rho_1=0,\rho_2=0,\rho_3=N_s$). 
The value of $\theta$ is given by a Gaussian random variable centered at 
zero with a standard deviation $\theta_T=2/\sqrt{N_s}$.
Since there is no incoherent pumping at $t>0$, $P_r$ is set to 0.

\begin{figure}
  \centering
  \includegraphics[width=8.5cm]{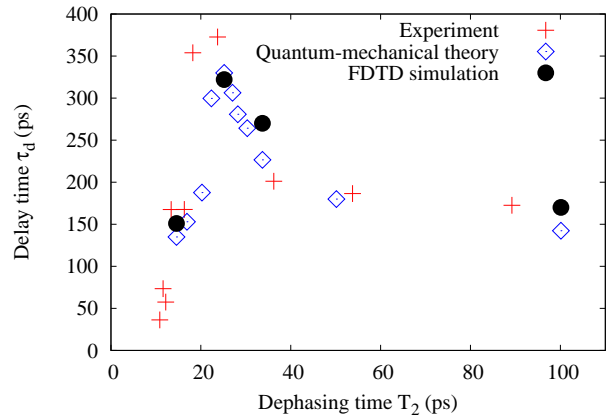} 
  \caption{\label{fig:fig2} (Color online)
    Comparison of delay times of emission pulse obtained by our numerical 
    simulation (black solid circles) with previous experimental data 
    (red crosses) and quantum-mechanical calculation results 
    (blue open diamonds) taken from ref. \cite{maki89}.
    Numerical delay times are obtained from the emission pulses averaged over 
    30 realizations. 
  }
\end{figure}

Figure \ref{fig:fig1} shows the output EM energy at a spatial grid point 
outside the system for four different values of the dephasing time $T_2$. 
When $T_2 = 100$ ps $ > \tau_d$, the cooperative emission characteristic of SF 
is clearly seen in Fig. \ref{fig:fig1}(a). 
The number of atoms that emit cooperatively is estimated to be 
$N_c = {8\pi c T_1 A}/{3\lambda^2 L} = 3.5 \times 10^8$ and is known as the 
Arecchi-Courtens cooperation number. 
Since $N_c < N = 3\times 10^9$, the SF oscillates in time, with the maximal 
emission intensity at $t \sim 170$ ps. 
This behavior agrees well with the previous result in \cite{maki89}.
For $T_2$ = 33.3 ps $< \tau_d$, there is enough dephasing to disturb the 
cooperative emission. 
The emitted pulse broadens and the time delay increases, as shown in 
Fig. \ref{fig:fig1}(b). 
For $T_2$ = 25 ps, a further damping of superfluorescence is seen in 
Fig. \ref{fig:fig1}(c). 
As $T_2$ decreases more, the pulse continues to broaden but the time delay 
begins to decrease. 
When $T_2$ reaches the critical value $\sqrt{\tau_r\tau_d}$ = 15.9 ps, 
the amount of dephasing is sufficient to prevent the occurrence of 
cooperative emission. 
No macroscopic dipole moment can build up and the atoms simply respond to 
the instantaneous value of the radiation field. 
Hence, SF is replaced by ASE. 
Figure \ref{fig:fig1}(d) plots the ASE pulse for $T_2=14.3$ ps. 
The time delay is almost immeasurably small 
and the emission intensity is very noisy. 
Figure \ref{fig:fig2} compares the delay times taken from our FDTD 
simulations to previous results obtained experimentally \cite{malcuit87} 
and by full quantum-mechanical theory of SF \cite{maki89}. 
The excellent agreement validates our FDTD-based numerical method. 
We emphasize that inclusion of the noise terms in (\ref{eq:theory}) is essential to obtain the correct variation of $\tau_d$ with $T_2$.  
As found in \cite{malcuit87}, the previous approach of modeling the initial fluctuations as random tipping angles of the Bloch vector and 
ignoring the noise at later times brings
about good agreement with experiment only when $T_2$ is large making the amplitude of the noise terms in (\ref{eq:theory}) small. 
As the dephasing rate increases, fluctuations can no longer be modeled simply as an initial noise.

\begin{figure}
  \centering
  \includegraphics[width=8.5cm]{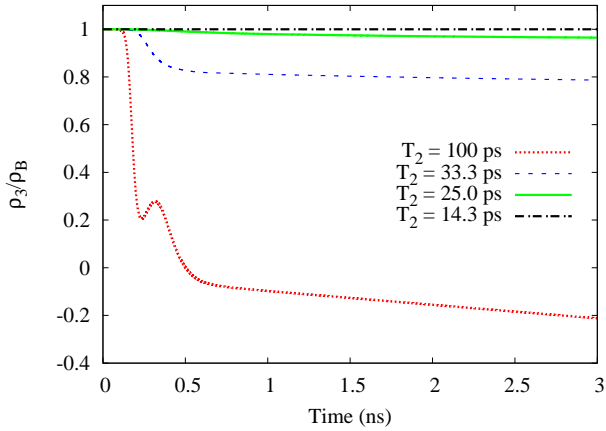}
  \caption{\label{fig:fig3} (Color online)
    The ratio $\rho_3/\rho_B$ as a function of time for $T_2$ = 100 ps 
    (red dotted line), 33.3 ps (blue dashed line), 25.0 ps (green solid line), 
    and 14.3 ps (black dash-dotted line).  
  }
\end{figure}

We have also studied the decoherence process. 
The amplitude of the Bloch vector 
$\rho_B \equiv \sqrt {\rho_1^2 + \rho_2^2 + \rho_3^2} = 
\sqrt{N_s^2 + 4\rho_{12}\rho_{21} - 4\rho_{22}\rho_{11} }$.
In the absence of decoherence, $\rho_{12}\rho_{21} = \rho_{22}\rho_{11}$, 
and $\rho_B = N_s$. 
The presence of decoherence decreases the off-diagonal terms of the density 
matrix, thus $\rho_{12}\rho_{21} < \rho_{22}\rho_{11}$ and 
$\rho_B < N_s$ \cite{bid01}. 
We estimate the degree of decoherence through the ratio $\rho_3/\rho_B$, 
which is plotted in Fig. \ref{fig:fig3} for four different values of $T_2$. 
Each curve is obtained by spatial average of $\rho_{3}$ and $\rho_{B}$ over 
the entire excitation region and then ensemble-average over 30 realizations.

When the dephasing time is large ($T_2 > \tau_d$), a macroscopic 
dipole moment is spontaneously formed. 
The enhanced radiative decay rate results in quick depletion of the population 
inversion $\rho_3$. 
Despite $T_2 \ll T_1$, the decay of $\rho_{1}$ and $\rho_{2}$ by dephasing is
overshadowed by the decay of $\rho_3$ by SF, leading to a rapid drop of 
$\rho_3/\rho_B$ in time.
This behavior is shown by the red dotted line in Fig. \ref{fig:fig3}.
The non-monotonic decay is caused by SF oscillations as can be seen in  
Fig. \ref{fig:fig1}(a).
The oscillatory SF is a result of the number of atoms being greater than the Arecchi-Courtens cooperation number ($N>N_c$). 
The intensity oscillation leads to an oscillation of population inversion which is 90 degree out of phase.
The local maximum of $\rho_3$ at $t=320$ ps (red dotted curve in \ref{fig:fig3}) occurs just before the second peak of intensity 
at $t=370 ps$ [Fig. \ref{fig:fig1}(a)].
As $T_2$ is reduced, the increased amount of decoherence frustrates the 
buildup of a macroscopic dipole moment and reduces the radiative decay rate. 
Consequently, the depletion of population inversion is slowed down.
It leads to a slower decay of $\rho_3 / \rho_B$ 
and the disappearance of damped oscillations.  
Finally when the dephasing time is small enough ($T_2 < \sqrt{\tau_r\tau_d}$), 
the system stays in a decoherent state, and $\rho_3 / \rho_B$ remains close to one 
for a very long time. 

\section{Conclusion}
We have developed a FDTD algorithm to incorporate stochastic noise in 
macroscopic systems into the Maxwell-Bloch equations. 
Such noise, resulting from atom-reservoir interactions, accompanies 
the dephasing of atomic polarization and decay of and pumping to the excited 
state population. 
We applied our algorithm to a numerical simulation of superfluorescence in a 
1D system.
The results are in good agreement with previous experimental and theoretical 
studies.
Although our simulations only include 
classical noise, 
nonclassical noise may be incorporated as well.
Since they consist of nonlinear terms \cite{drum91}, 
the incorporation of nonclassical fluctuations to the FDTD algorithm may be numerically challenging. 
Given the rapid progress in development of various numerical methods of including nonlinearity in the Maxwell-Bloch equations \cite{besse,bourg}, 
we are optimistic that the quantum noise terms may be successfully integrated into our method.
Therefore, our FDTD-based model can be used for numerical studies of 
light-matter interaction and transient processes in complex systems without 
prior knowledge of modes.

\section*{Acknowledgments}
This work is supported by the NSF under the Grant Nos. DMR-0814025 and 
DMR-0808937. 
The authors acknowledge Profs. Prem Kumar,  Allen Taflove and Changqi Cao for 
stimulating discussions. 
The authors also thank the Yale University Biomedical High Performance
Computing Center and NIH Grant RR19895 which funded the instrumentation.



\end{document}